\PassOptionsToPackage{unicode}{hyperref}
\PassOptionsToPackage{hyphens}{url}
\documentclass[]{article}

\usepackage{arxiv}

\usepackage[utf8]{inputenc} % allow utf-8 input
\usepackage[T1]{fontenc}    % use 8-bit T1 fonts
\usepackage{hyperref}       % hyperlinks
\usepackage{url}            % simple URL typesetting
\usepackage{booktabs}       % professional-quality tables
\usepackage{amsfonts}       % blackboard math symbols
\usepackage{nicefrac}       % compact symbols for 1/2, etc.
\usepackage{microtype}      % microtypography
\usepackage{lipsum}
\usepackage{graphicx}
\usepackage{xcolor}
\hypersetup{
    colorlinks,
    linkcolor={red!50!black},
    citecolor={blue!50!black},
    urlcolor={blue!80!black}
}

\graphicspath{ {./images/} }

\title{Metatheory.jl: Fast and Elegant Algebraic Computation in Julia
with Extensible Equality Saturation}
\author{  Alessandro Cheli \\
Department of Computer Science\\
 University of Pisa\\
 Pisa, Italy, 56127 \\
 \texttt{a.cheli6@studenti.unipi.it} }
\date{11 February 2021}

\begin{document}

\maketitle

\begin{abstract} 
We introduce Metatheory.jl: a lightweight and performant general purpose
symbolics and metaprogramming framework meant to simplify the act of writing
complex Julia metaprograms and to significantly enhance Julia with a native term
rewriting system, based on state-of-the-art equality saturation techniques, and
a dynamic first class AST pattern matching system that is dynamically composable
in an algebraic fashion, taking full advantage of the language's powerful reflection
capabilities. Our contribution allows to perform general purpose symbolic
mathematics, manipulation, optimization, synthesis or analysis of syntactically
valid Julia expressions with a clean and concise programming interface, both
during compilation or execution of programs.
\end{abstract}

\hypertarget{statement-of-need}{%
\section{Statement of Need}\label{statement-of-need}}

The Julia programming language is a fresh approach to technical
computing \cite{bezanson2017julia}, disrupting the popular conviction
that a programming language cannot be high level, easy to learn, and
performant at the same time. One of the most practical features of Julia
is the excellent metaprogramming and macro system, allowing for
\emph{homoiconicity}: programmatic generation and manipulation of
expressions as first-class values, a well-known paradigm similar to
several LISP idioms such as Scheme.

We introduce Metatheory.jl: a general purpose metaprogramming and
algebraic computation library for the Julia programming language,
designed to take advantage of the powerful reflection capabilities to
bridge the gap between symbolic mathematics, abstract interpretation,
equational reasoning, optimization, composable compiler transforms, and
advanced homoiconic pattern matching features. Intuitively,
Metatheory.jl transforms Julia expressions in other Julia expressions at
both compile and run time. This allows Metatheory.jl users to perform
customized and composable compiler optimization specifically tailored to
single, arbitrary Julia packages. Our library provides a simple,
algebraically composable interface to help scientists in implementing
and reasoning about all kinds of formal systems, by defining concise
rewriting rules as syntactically valid Julia code.

\hypertarget{summary}{%
\section{Summary}\label{summary}}

Theories can then be executed throught two, highly composable, rewriting
backends. The first one, is based just on standard rewriting built on
top of the pattern matcher developed in \cite{matchcore}. Such approach
suffers of the usual problems of rewriting systems: even trivial
equational rules such as commutativity may lead to non-terminating
systems and thus need to be adjusted by some sort of structuring or
rewriting order (that is known to require extensive user reasoning).

The other back-end for Metatheory.jl, the core of our contribution, is
designed to not require the user to reason about rewriting order by
relying on state-of-the-art techniques equality saturation on
\emph{e-graphs}, adapted from the \texttt{egg} rust library \cite{egg}.
Provided with a theory of equational rewriting rules, defined in pure
Julia, \emph{e-graphs} compactly represent many equivalent programs.
Saturation iteratively executes an e-graph specific pattern matcher to
efficiently compute (and analyze) all possible equivalent expressions
contained in the e-graph congruence closure. This latter back-end is
suitable for partial evaluators, symbolic mathematics, static analysis,
theorem proving and superoptimizers.

\begin{figure}
\centering
\includegraphics[width=\textwidth]{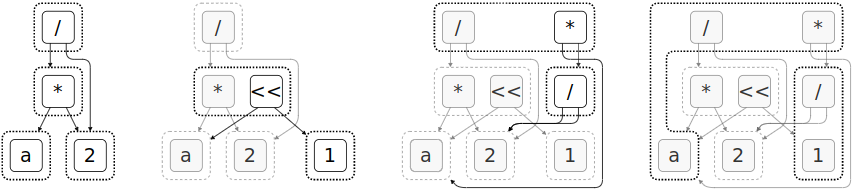}
\caption{These four e-graphs represent the process of equality
saturation, adding many equivalent ways to write \((a \times 2) / 2\)
after each iteration. Credits to Max Willsey (@egg).\label{fig:egggg}}
\end{figure}

The original \texttt{egg} library \cite{egg} is known to be the first
implementation of generic and extensible e-graphs \cite{nelson1980fast},
the contributions of \texttt{egg} include novel amortized algorithms for
fast and efficient equivalence saturation and analysis. Differently from
the original rust implementation of \texttt{egg}, which handles
expressions defined as rust strings and data structures, our system
directly manipulates homoiconic Julia expressions, and can therefore
fully leverage the Julia subtyping mechanism \cite{zappa2018julia},
allowing programmers to build expressions containing not only symbols
but all kinds of Julia values. This permits rewriting and analyses to be
efficiently based on runtime data contained in expressions. Most
importantly, users can --and are encouraged to-- include type assertions
in the left hand of rewriting rules in theories.

One of the project goals of Metatheory, beyond being to be easy to use
and composable, is to be fast and efficient: both the first-class
pattern matching system and the generation of e-graph analyses from
theories rely on RuntimeGeneratedFunctions.jl \cite{rgf}, generating
callable functions at runtime that efficiently bypass Julia's world age
problem \cite{belyakova2020world} with the full performance of a
standard Julia anonymous function.

\hypertarget{analyses-and-extraction}{%
\subsection{Analyses and Extraction}\label{analyses-and-extraction}}

With Metatheory.jl, modeling analyses and conditional/dynamic rewrites
is straightforward: it is possible to check conditions on runtime values
or to read and write from external data structures during rewriting. The
analysis mechanism described in egg \cite{egg} and re-implemented in our
contribution lets users define ways to compute additional analysis
metadata from an arbitrary semi-lattice domain, such as costs of nodes
or logical statements attached to terms. Other than for inspection,
analysis data can be used to modify expressions in the e-graph both
during rewriting steps and after e-graph saturation.

Therefore using the equality saturation (e-graph) backend, extraction
can be performed as an on-the-fly e-graph analysis or after saturation.
Users can define their own, or choose between a variety of predefined
cost functions for automatically extracting the most fitting expressions
from the congruence closure represented by an e-graph.

\hypertarget{conclusion}{%
\section{Conclusion}\label{conclusion}}

Many applications of equality saturation have been recently published,
tailoring advanced optimization tasks. Herbie
\cite{panchekha2015automatically} is a tool for automatically improving
the precision of floating point expressions, which recently switched to
\texttt{egg} as the core rewriting backend. In \cite{yang2021equality},
authors used \texttt{egg} to superoptimize tensor signal flow graphs
describing neural networks. However, Herbie requires interoperation and
conversion of expressions between different languages and libraries.
Implementing similar case studies in pure Julia would make valid
research contributions on their own. We are confident that a
well-integrated and homoiconic equality saturation engine in pure Julia
will permit exploration of many new metaprogramming applications, and
allow them to be implemented in an elegant, performant and concise way.
Code for Metatheory.jl is available in \cite{metatheory}, or at this
link \url{https://github.com/0x0f0f0f/Metatheory.jl}.

\hypertarget{acknowledgements}{%
\section{Acknowledgements}\label{acknowledgements}}

We acknowledge Max Willsey and contributors for their work on the
original \texttt{egg} library \cite{egg}, Christopher Rackauckas and
Christopher Foster for their efforts in developing
RuntimeGeneratedFunctions \cite{rgf}, Taine Zhao for developing MLStyle
\cite{mlstyle} and MatchCore \cite{matchcore}, and Philip Zucker for his
original idea of implementing E-Graphs in Julia \cite{philzuck1,
philzuck2} and support during the development of the project. Special
thanks to Filippo Bonchi for a friendly review of a preliminary version
of this article.

\bibliographystyle{apalike}  
\bibliography{paper}

\end{document}